\documentclass[prl,showpacs,preprintnumbers,amsmath,amssymb,superscriptaddress,twocolumn,floatfix]{revtex4}
\usepackage{graphicx}% Include figure files

\begin{document}
\newcommand{\bNMR}{$\beta${\sc -nmr}}
\newcommand{\bnmr}{$\beta${\sc -nmr}}
\newcommand{\bNQR}{$\beta${\sc -nqr}}
\newcommand{\nmr}{{\sc nmr}}
\newcommand{\nqr}{{\sc nqr}}
\newcommand{\ie}{i.e.,}
\newcommand{\eg}{{\it e.g.}}
\newcommand{\etc}{{\it etc.}}
\newcommand{\etal}{{\it et al.}}
\newcommand{\ibid}{{\it ibid.}}
\newcommand{\dC}{$^{\circ}$C}
\newcommand{\Fig}[1]{{\mbox{Fig.~#1}}}
\newcommand{\eli}{$^{8}$Li}
\newcommand{\elip}{$^{8}$Li$^{+}$}
\newcommand{\elin}{$^{8}$Li$^{0}$}
\newcommand{\lip}{Li$^{+}$}
\newcommand{\ks}{\mbox{\it K}}
\newcommand{\xq}{\mbox{$T_{1}^{-1}$}}
\newcommand{\kr}{\mbox{${\mathcal K}$}}
\newcommand{\nuq}{\mbox{$\nu_{\it Q}$}}
%\twocolumn[
%\hsize\textwidth\columnwidth\hsize\csname@twocolumnfalse\endcsname

\title{$\beta$-NMR of isolated Lithium in nearly ferromagnetic Palladium}
\author{T.J.~Parolin}
\affiliation{Department of Chemistry, University of British Columbia, Vancouver, BC, V6T 1Z1, Canada}
\author{Z.~Salman}
\affiliation{TRIUMF, 4004 Wesbrook Mall, Vancouver, BC, V6T 2A3, Canada}
\author{J.~Chakhalian}
\altaffiliation[Former address: ]{Max-Plank-Institut f\"{u}r Festk\"{o}rperforschung, \mbox{D-70569}, Stuttgart, Germany.}
\affiliation{Department of Physics, University of Arkansas, Fayetteville, AR, 72701, USA.}
\author{Q.~Song}
\affiliation{Department of Physics and Astronomy, University of British Columbia, Vancouver, BC, V6T 1Z1, Canada}
\author{K.H.~Chow}
\affiliation{Department of Physics, University of Alberta, Edmonton, AB, T6G 2G7, Canada}
\author{M.D.~Hossain}
\affiliation{Department of Physics and Astronomy, University of British Columbia, Vancouver, BC, V6T 1Z1, Canada}
\author{T.A.~Keeler}
\affiliation{Department of Physics and Astronomy, University of British Columbia, Vancouver, BC, V6T 1Z1, Canada}
\author{R.F.~Kiefl}
\affiliation{TRIUMF, 4004 Wesbrook Mall, Vancouver, BC, V6T 2A3, Canada}
\affiliation{Department of Physics and Astronomy, University of British Columbia, Vancouver, BC, V6T 1Z1, Canada}
\affiliation{Canadian Institute for Advanced Research}
\author{S.R.~Kreitzman}
\affiliation{TRIUMF, 4004 Wesbrook Mall, Vancouver, BC, V6T 2A3, Canada}
\author{C.D.P.~Levy}
\affiliation{TRIUMF, 4004 Wesbrook Mall, Vancouver, BC, V6T 2A3, Canada}
\author{R.I.~Miller}
\affiliation{TRIUMF, 4004 Wesbrook Mall, Vancouver, BC, V6T 2A3, Canada}
\author{G.D.~Morris}
\affiliation{TRIUMF, 4004 Wesbrook Mall, Vancouver, BC, V6T 2A3, Canada}
\author{M.R.~Pearson}
\affiliation{TRIUMF, 4004 Wesbrook Mall, Vancouver, BC, V6T 2A3, Canada}
\author{H.~Saadaoui}
\affiliation{Department of Physics and Astronomy, University of British Columbia, Vancouver, BC, V6T 1Z1, Canada}
\author{D.~Wang}
\affiliation{Department of Physics and Astronomy, University of British Columbia, Vancouver, BC, V6T 1Z1, Canada}
\author{W.A.~MacFarlane}
\affiliation{Department of Chemistry, University of British Columbia, Vancouver, BC, V6T 1Z1, Canada}

\date{November 23, 2006}

\begin{abstract}
The temperature dependence of the frequency shift and spin-lattice relaxation rate of isolated, nonmagnetic $^8$Li impurities implanted in a nearly ferromagnetic host (Pd) are measured by means of $\beta$-detected nuclear magnetic resonance (\bnmr). The shift is negative, very large and increases monotonically with decreasing $T$ in proportion to the bulk susceptibility of Pd for $T > T^{*}\approx 100$ K. Below $T^*$, an additional shift occurs which we attribute to the response of Pd to the defect. The relaxation rate is much slower than expected for the large shift and is linear with $T$ below $T^*$, showing no sign of additional relaxation mechanisms associated with the defect.
\end{abstract}


\maketitle

Elemental metallic\cite{metal} Palladium is on the verge of ferromagnetism as evidenced by its large temperature-dependent paramagnetic susceptibility $\chi$.
The chemical and structural simplicity of Pd makes it a particularly appealing example of a nearly ferromagnetic (NF) metal.
Efforts to understand NF metals have led to significant theoretical progress, such as the advent of the spin fluctuation model (SFM) that has recently been adapted to the nearly {\it antiferromagnetic} doped cuprates\cite{sf}.
While the SFM has been successfully applied to Pd, detailed calculations of quantities such as $\chi$ still present a challenge\cite{newth}.
Many recent attempts to understand the rich variety of unconventional properties of nearly magnetic materials are based on the paradigm of quantum criticality, where proximity to a zero temperature quantum critical point (QCP) controls the material properties over a wide range of the phase diagram.
In this context, Pd can be tuned towards an itinerant ferromagnetic groundstate either
by introducing dilute magnetic impurities\cite{dope} or by expanding its lattice in an epitaxial heterostructure\cite{hetero}.
For example, refined studies of $Pd$Ni indicate the QCP occurs at a Ni concentration of only 2.5\%\cite{qct},
thus pure Pd is in the realm of influence of this ferromagnetic QCP, although it is clearly a Fermi liquid\cite{metal}.
The defect response of a metal near a QCP is highly unconventional and not yet understood.
For example, it has been predicted that droplets of local order (interacting with the quantum critical environment)
will be nucleated by a pointlike defect\cite{millis}.

In this Letter, we present an \nmr\ study of an isolated atomic defect in pure Pd. In particular, we
implant spin polarized radioactive $^8$Li and detect the \nmr\ via the parity violating weak $\beta$-decay ($\beta$-\nmr )
in a Au(10 nm)/Pd(100 nm)/SrTiO$_3$ heterostructure and in a thin Pd foil, yielding
a local measure of the magnetic character of the Li defect through the \nmr\ shift and spin lattice relaxation rate.
Results in the film agree with measurements in the bulk foil but are better resolved and have
an {\it in situ} reference from the Au layer, allowing an accurate measure of the shift in Pd.
We find a very large, strongly temperature-dependent Knight shift $K$ which follows
the host $\chi(T)$ down to $T^* \approx 100$ K, with a deviation below $T^*$ that
we attribute to the response of Pd to the defect.
In contrast, the spin-lattice relaxation rate $T_{1}^{-1}$ is remarkably slow and shows a simple
Korringa ($\propto T$) behaviour below $T^*$.

\nmr\ is a powerful technique; one of the few that can reveal both the average magnetic behaviour {\it and} its microscopic inhomogeneity.  Application of conventional \nmr\ to thin film heterostructures is limited by sensitivity; however, \bnmr\ with a
low energy beam of radioactive \elip\ can be used in this case\cite{agprl,tak}.
In the \nmr\ of metals, the relative frequency shift of the resonance, $\delta=(\nu - \nu_{r}) / \nu_{r}$, where $\nu_r$ is the reference frequency, is composed of two contributions $\delta = K + K^{orb}$: the Knight shift ($K$) resulting from Fermi contact coupling to the Pauli spin susceptibility of the conduction electrons, and the temperature-independent orbital (chemical) shift $K^{orb}$.
In conventional metals the spin relaxation rate is linear with temperature, following the Korringa Law, $(T_{1}T)^{-1} \propto K^{2} $. In contrast in NF metals, low frequency, long wavelength spin fluctuations modify the Korringa Law, strongly enhancing the
proportionality constant at low temperature, and causing a high temperature deviation such that $(T_{1}T)^{-1}$ is instead proportional to $\chi$\cite{m-u}, as seen \eg\ in TiBe$_2$\cite{alloul}.
Conventional \nmr\ of $^{105}$Pd\cite{pdnmr,taki} showed that $K(T)$ follows the bulk $\chi(T)$,
and the observed \xq\ varies linearly below 150 K.

The present experiments were carried out on the polarized low energy beamline at TRIUMF's ISAC facility\cite{agprl,zssto}.  The \elip\ probe nuclei ($I= 2$, $\tau= 1.2$ s,  $^8\gamma= 6.3015$ MHz/T) are implanted at an energy controlled by electrostatic
deceleration. The observed asymmetry of the $\beta$-electron count rate ($A$) is proportional to the \eli\ nuclear spin polarization.  For resonance measurements, we use a continuous beam of $\sim 10^6$ ions/s focussed to a 3 mm diameter beamspot; measure the time-integrated $A$ for 1 s; then step the frequency of the small transverse radiofrequency (RF) magnetic field $H_{1}$.
The spin-lattice relaxation rate $T_1^{-1}$ is measured by pulsing the incident ion beam (4 s pulsewidth) and monitoring the time dependence of $A$ with the RF off.  $A$(t) is fit to an exponential form during {\it and} after each beam pulse\cite{sto06}.  All the data were taken in a static external magnetic field $H_0$= 4.1 T\cite{fn2}.
The high sensitivity of the nuclear detection allows us to work in the extremely dilute limit (the \eli\ concentration is about 1 in $10^{11}$), yielding the properties of isolated Li. 

The resonances were measured in a 100 nm film grown via e-beam deposition from a 99.99\% source (Goodfellow)
onto an epitaxially polished $\langle 100\rangle$ crystalline SrTiO$_3$ (STO) substrate at 60--80\dC.
The growth rate was 0.5\AA/s under a background pressure of $\sim$10$^{-8}$ Torr.
It was subsequently capped {\it in situ} with 10 nm of Au.
The Pd layer was found to be highly oriented in the $\langle 111\rangle$ direction by x-ray diffraction.
An implantation energy of 11 keV was used as Monte Carlo simulations of the \elip\ stopping profile (using TRIM.SP\cite{trim}) indicated that this would maximize the number of ions stopping in the Pd.
The \xq\ data was collected in a 12.5 $\mu$m thick foil of 99.95\% Pd (Alfa Aesar).

\begin{figure}
\centerline{\includegraphics[width=\columnwidth,angle=0]{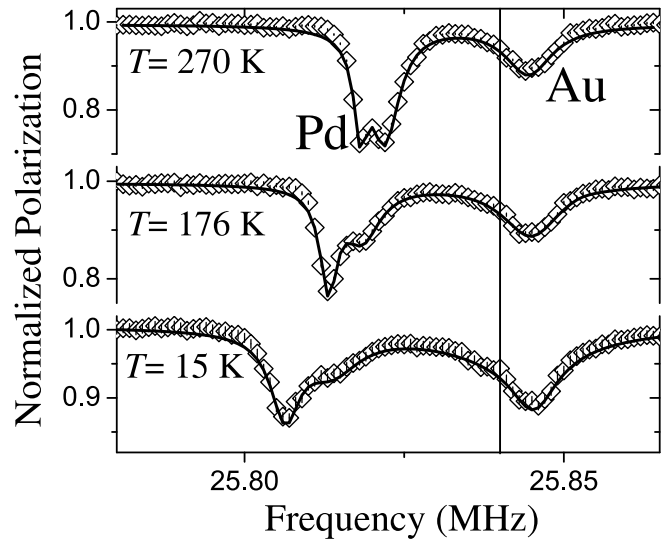}}
\caption{
\bnmr\ spectra of \eli\ in the Au/Pd/STO film fit to three Lorentzians.  The vertical line is the reference frequency $\nu_r$.
The Pd signal is fit to two resonances corresponding to different stopping sites.
}
\label{1}
\end{figure}

Three representative spectra are displayed in Fig.\ 1. Resonances from \eli\ in both Au and Pd were identified by comparison with previous measurements in Au\cite{wam03} and in Pd foil\cite{usr}, as well as the dependence of their relative amplitudes on
implantation energy. The shift of the \eli\ resonance in Au is +60 ppm relative to the resonance in the cubic insulator MgO\cite{au06}.
We use this to infer the reference frequency (in MgO) $\nu_r= 25.842$ MHz (vertical line in Fig.\ 1) used throughout.
Similar to previous measurements in unoriented polycrystalline Pd foil, the shift is large and negative, becoming more negative with decreasing $T$, confirming it is (a) intrinsic to Li in Pd and (b) predominantly isotropic. At 270 K, the Pd signal is clearly split into two resonances of near equal amplitude. The splitting is magnetic, rather than quadrupolar, implying two distinct Li sites of cubic
symmetry. As $T$ is reduced, the less shifted resonance diminishes in amplitude as the more shifted line becomes predominant.
Previous \bnmr\ studies of \eli\ in Ag\cite{agprl} and Au\cite{wam03} show that Li typically occupies two distinct cubic sites
in FCC metals: the interstitial octahedral ($O$) and substitutional ($S$) sites, each with a distinct shift.
It is likely that the observed signals are due to \eli\ stopping in similar sites in isostructural Pd, but the
two shifts are surprisingly similar compared to Ag and Au, where they differ by a factor of $\sim 2$.
In Ag and Au the more shifted $O$ site resonance is metastable, disappearing above $\sim170$ K as Li makes a thermally activated site
change. The current data suggest a similar site change occurs in Pd above 250 K. More measurements are required to confirm this
site assignment. The two resonances are, however, very close, and they track one another as a function of $T$, so
we consider only the average shift hereafter.

\begin{figure}
\centerline{\includegraphics[width=\columnwidth,angle=0]{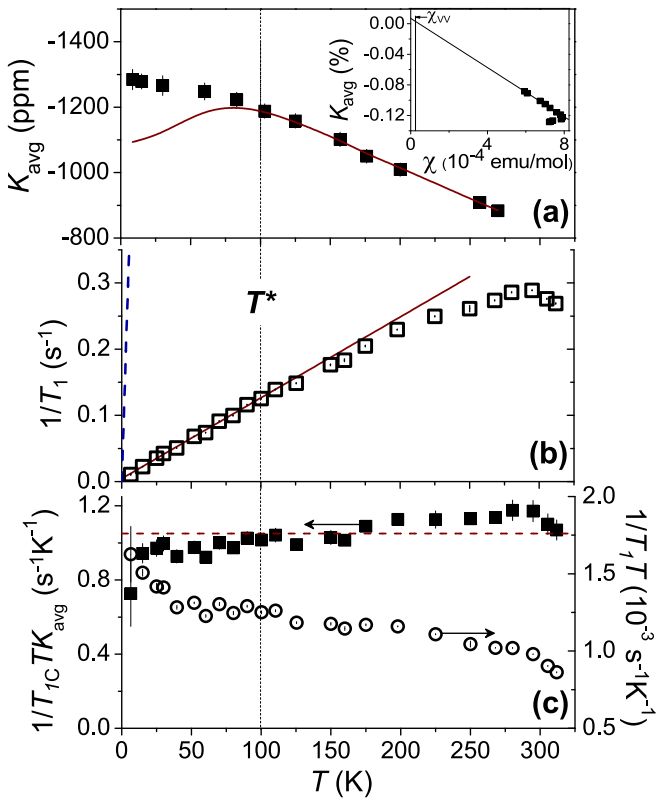}}
\caption{
(color online).  (a) Temperature dependence of the average shift of \eli\ in Pd. (inset)  $K_{\mathrm avg}$ vs. $\chi$. $K(T)$ obtained from a fit to the linear region is represented by the solid red line in the main panel.  (b) Temperature dependence of the \xq\ of \elip\ in Pd foil.  The red line is a fit to the low $T$ data and the dashed blue line is the Korringa-predicted \xq\ for $K_{\mathrm avg}$= -885 ppm.  (c) The product $T_{1C}TK_{\mathrm avg}$ is nearly independent of $T$ (squares; left scale).  Some $K_{\mathrm avg}$ have been interpolated/extrapolated from the data in (a).  $(T_{1}T)^{-1}$ is also shown for comparison (circles; right scale).
}
\label{2}
\end{figure}

A plot of $\delta_{\mathrm avg}(T)$ vs.\ $\chi(T)$ of bulk Pd\cite{pdnmr} [inset Fig.\ 2(a)] shows a clear proportionality for $T\geq T^*$.  Extrapolating this line to the zero of $\chi$ (or to the small van Vleck $\chi$), we find $K^{orb}$ is quite small ($40\pm240$ ppm) as one would expect for a light atom. This estimate of $K^{orb}$ is rather uncertain due to the large extrapolation, so rather than subtracting it from $\delta$ to obtain $K$, we instead assume $K \approx \delta$.
From the slope of this fit we extract the hyperfine coupling $A_{hf}$ using $K_{\mathrm avg}(T)=(A_{hf}Z/N_{A}\mu_{B})\chi(T)$; where $N_A$ is Avogadro's number, $\mu_B$ the Bohr magneton, and $Z$ the co-ordination of the site assumed to be 6 ($O$-site), and obtain $A_{hf} =-1.50(2)$ kG/$\mu_{B}$, a substantially smaller magnitude than for Li in Ag\cite{agprl} or Au\cite{au06}.
For Pd, $\chi(T)$ exhibits a characteristic maximum at $T^*$. Within the SFM, this is due to thermal excitation of spin fluctuations
producing effective magnetic moments. Once the temperature-induced moments reach a saturation amplitude ($\sim T^*$),
they are then depolarized by higher energy thermal fluctuations, giving rise to the Curie-Weiss high $T$ regime
which is well described by Moriya's self-consistent renormalization approach\cite{moriya-book}.
For Li in Pd, below $T^*$, the shift diverges from $\chi(T)$ of Pd (solid line in Fig.\ 2(a)).

We turn now to the spin-lattice relaxation.  Previous results for \eli\ in Ag demonstrated that the implanted ions are coupled 
to the conduction electrons and obey the Korringa Law quite closely\cite{agprl}.
The Pd film was not suitable for $T_1$ measurements as a fraction of the Li always stops in the Au overlayer. To avoid such contamination, we instead measured $T_1$ in a Pd foil which exhibits similar resonances\cite{usr}. Representative relaxation data and fits are shown in Fig.\ 3. The relaxation is predominantly single exponential, with a small ($<10$\%) background signal from backscattered Li.
The relaxation rates \xq\ are shown in Fig.\ 2(b). A linear fit below $T^*$ yields \xq$=1.22(2)\times10^{-3}$s$^{-1}$K$^{-1}T+4.5(9)\times10^{-3}$s$^{-1}$. To extract the purely Korringa relaxation $T_{1C}^{-1}$, we subtract the small $T$ independent term from the observed rate .
Above $T^*$, $T_1^{-1}(T)$ is clearly sublinear. 
This tendency to saturation at high temperatures
is qualitatively similar to the temperature independent relaxation found in local moment 
paramagnets; however, a correct explanation requires modelling the spin fluctuation spectrum\cite{taki}.
The nearly temperature independent $(T_{1C}TK_{\mathrm avg})^{-1}$ [Fig. 2(c)] is consistent
with NF behaviour and with the Pd $T_1$\cite{taki}, while the maximum in
\xq\ near room temperature is not. Rather, it is likely due to a Li site change to a high
temperature site characterized by a smaller shift and Korringa slope, as has been seen in other FCC metals\cite{au06,cu06}.

\begin{figure}
\centerline{\includegraphics[width=\columnwidth,angle=0]{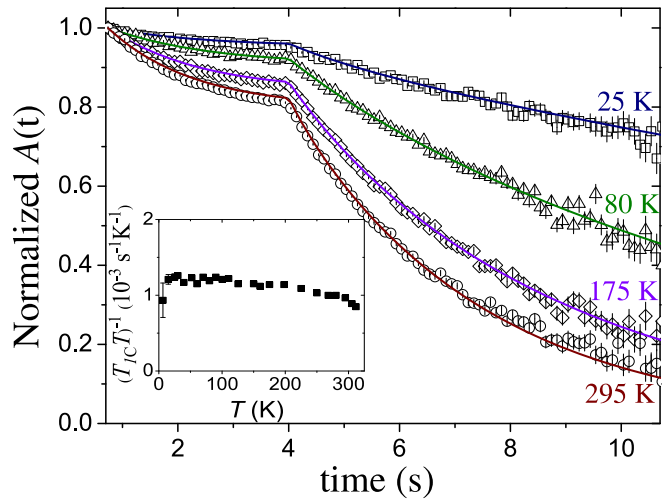}}
%\vskip 0.2cm
\caption{
(color online).  Normalized spin relaxation data and fits. (inset) Temperature dependence of $(T_{1C}T)^{-1}$.
}
\label{3}
\end{figure}

We begin by discussing the origin of the large shift $K$.
An impurity nucleus in a host metal is generally coupled to the conduction band
via a transferred hyperfine coupling with its near neighbors. This interaction depends on the overlap of the impurity atomic orbitals with the neighboring conduction band orbitals, \ie\ on the bonding of the impurity to the host. 
In Ag, Au and other simple metals this typically results in a small, positive, temperature-independent
$K$ for Li. In contrast in Pd, we observe a shift which is negative, temperature-dependent and about 10 times larger.
The first point is in fact expected for a coupling arising from ``$s-d$ hybridization'' between the Li $2s$ and Pd $4d$ orbitals
\cite{imp-calc}, while the latter two points result from the very large $T$ dependent $\chi$ of Pd. A recent calculation predicts
$K$ for Li in Pd within a factor of 2 of the observed value\cite{min}. The shift from such a coupling is isotropic,
as observed. 

The deviation of the Li shift $K$ from $\chi$ below $T^*$ reflects a local modification of $\chi$ due to
the nonmagnetic defect. In contrast, in Pd \nmr \cite{pdnmr,taki}, the $K-\chi$ scaling is maintained to low temperature.
Identical behavior for $K$ of Li in Pd foil\cite{usr} confirms the deviation is intrinsic.
Indeed such a response is similar to the evolution of $\chi$ in dilute alloys of $Pd$Ag\cite{pdrhag},
where 2.5\% Ag is sufficient to eliminate the maximum in $\chi(T)$. In such alloys, despite their dilution, it is difficult to
rule out $4d$ band-filling as the origin. In contrast, here the Li is in the dilute limit.  
We note similar behavior appears in $K(T)$ of the implanted $\mu^+$ in Pd\cite{gyg}.
These similarities suggest a common origin which is quite insensitive to the details of the nonmagnetic impurity potential.
Such an explanation has in fact been suggested on theoretical grounds\cite{chi-imp}. It would be interesting to
test this theory with a detailed calculation of the \lip\ defect in Pd.

A major difficulty in interpreting Pd \nmr\ is decomposing $K$ and \xq\ into orbital and spin contributions\cite{taki}.
For Li, we do not expect orbital effects since the atomic electrons are in a state of zero orbital angular momentum.
This is confirmed by the small size of $K^{orb}$. Similarly, we expect the orbital contribution to \xq\ to be negligible\cite{orbLN}.
In both Ag and Au, the Li relaxation rate follows the Korringa Law: $(T_{1}T)^{-1}$ is within a factor of 2 of values predicted from the shifts. If Pd were similar, the Korringa \xq s should be much larger [\eg\ dashed line Fig.\ 2(b)], and
any additional orbital relaxation would only enhance this discrepancy. Millis {\it et al.} predict that a defect in an itinerant
NF system will nucleate a local order droplet with a magnetic fluctuation spectrum that is locally highly suppressed at low frequency\cite{millis}. While pure Pd may not be close enough to the QCP for this theory to apply in detail, it is qualitatively consistent with our observation that below $T^*$, where the shift contains a defect contribution, we obtain only very slow
$T$-linear relaxation. As the impurity contribution to the total shift is relatively small ($< 20$\%), the
deviation in {\bf $(T_{1C}TK_{\mathrm avg})^{-1}$} below $T^*$ is only apparent as a small slope at low temperature.

In conclusion, the shift and spin-lattice relaxation rate of highly dilute nonmagnetic \eli\ impurities implanted in NF Pd were
measured via \bnmr.  The shifts are isotropic, highly temperature-dependent, and scale with the host $\chi$ above a characteristic $T^*$, below which they continue to increase until saturating at the lowest $T$.  The anomalous additional shift at low $T$ is attributed to a defect response of the Pd. The Korringa relaxation is much slower than expected for the large value of the shift. The product
$(T_{1C}TK_{\mathrm avg})^{-1}$ is approximately $T$-independent as expected for a NF metal. It will be interesting to
follow the evolution of these properties in both thinner films (where finite size effects are expected to play an important role) and in alloy films and heterostructures closer to quantum criticality.

We acknowledge the technical assistance of R.\ Abasalti, B.\ Hitti, D.\ Arseneau, S.\ Daviel, H.-U. Habermeier and M. Xu,
and funding from NSERC Canada.  TRIUMF is funded in part by the NRC.


\begin{thebibliography}{99}
\bibitem{qct} Here the quantum critical point is the concentration $x$ where the ferromagnetic $T_C$ is just suppressed to $0$, see M.~Nicklas \etal, Phys. Rev. Lett. {\bf 82}, 4268 (1999).
\bibitem{fn2}For thin films the demagnetizing factor $\sim 1$, thus the internal field is equal to the applied field.
\end{thebibliography}
\end{document}